\documentstyle[epsf,aps,12pt]{revtex}  

\begin{document}
\draft
\title{Role of isospin dependent mean field in 
pion production in heavy ion reactions
}
\author{V.S. Uma Maheswari, C. Fuchs, Amand Faessler, 
Z.S. Wang and D.S. Kosov
}
\address{Institut f\"ur Theoretische Physik der Universit\"at T\"ubingen,
\\ Auf der Morgenstelle 14, D-72076 T\"ubingen, Germany
}
\maketitle  
\begin{abstract}
The importance of a isospin dependent nuclear mean field (IDMF) 
in regard to the pion production mechanism is studied 
for the reaction $Au+Au$ at 1 GeV/nucleon
using the Quantum Molecular Dynamics (QMD) model. 
In particular, the effect of the IDMF on pion spectra
and the charged pion ratio are analyzed. 
It is found that the inclusion
of a IDMF considerably suppresses the low$-p_t$ pions, thus, 
leading to a better agreement with the data on pion spectra.
Moreover, the rapidity distribution
of the charged pion ratio appears to be sensitive to the isospin 
dependence of the nuclear mean field.
\end{abstract}
\newpage
\section{Introduction}
The investigation of particle production is a well established
method to probe the hot and dense nuclear matter which is produced
in collisions between heavy nuclei. 
The collision dynamics is rather fairly understood within transport
model approaches. At incident energies around 1 A.GeV one finds that 
the pion production is the dominant channel by which the hot, compressed
matter de-excites. Pionic observables such as yields, spectra and flow 
have been measured at GSI (FOPI \cite{pelt97}, KaoS \cite{bril93,munt95}) 
 and LBL \cite{naga81,broc84,kint97}.
One, however, finds that transport models such as QMD and 
BUU in general overestimate the abundance of low energy pions as well as the
charged pions multiplicities \cite{pelt97,teis97,uma97}.
In addition, this discrepancy is most prominent in
heavy colliding systems like $Au$, as compared to light systems
such as $Ni$ \cite{pelt97a}.
One characteristic aspect of such heavy systems is the high 
neutron-proton asymmetry effect. Essentially due to this
asymmetry factor $I={N-Z\over A}$, in the case of heavy
systems, one generally finds for the total pion yields 
 $N(\pi^-) > N(\pi^0) > N(\pi^+)$. On the other hand, 
for light systems one expects the pion abundances to be more 
or less equally distributed, i.e. 
$N(\pi^-) \simeq N(\pi^0) \simeq N(\pi^+)$. 
This observed charge dependence of the pion yields is
mainly due to the isospin dependence of the pion production
cross sections which can be well understood within the
isobar model \cite{stoc86,verw82,eric88} 
as most of the pions produced at intermediate energies 
originate from $\Delta-$decays.

Normally, although a isospin dependence is duly included in the
baryon-baryon collision process, these baryons comprising
nucleons, $\Delta's$ and $N^{*'}s$ propagate in an isospin
independent mean field.
As most of the pions, particularly the low energy ones, 
result from $\Delta-$decays the general nature of the
production/re-absorption and the propagation of $\Delta-$resonances
are important. The inclusion of a isospin dependent mean field (IDMF) 
will certainly affect the
reaction dynamics of nucleons and resonances leading to 
possible differences in pion yields. 
This is due to the fact that the various charges of nucleons 
and resonances will be affected in a different way 
by the isospin dependent mean field during their propagation.
This isospin dependence makes neutrons more energetic than protons. 
Thus, with other properties remaining the same, one expects
that with inclusion of a IDMF the average number $N_{<nn>}$ of $n$--$n$ 
collisions will get somewhat reduced while the average number $N_{<pp>}$
of $p$--$p$ collisions will get slightly enhanced.
At low incident energies, it is found \cite{li96,li97} that the IDMF 
strongly affects observables such as the collective flow and 
the pre-equilibrium $p/n$ ratio. Therefore, it is of 
general interest to analyze the role of the IDMF in the
context of pion production in heavy ion reactions at SIS energies.

In view of this we make our first exploratory study on the 
role of a isospin dependent mean field on the pionic observables
such as $p_t-$spectra and the $\pi^-/\pi^+$ ratio.

\section{Isospin dependence of pion production }

To analyse the effect of the isospin dependence on the pion
production process we make use of the Quantum Molecular
Dynamics (QMD) model \cite{aich91} which is well described in our earlier 
works \cite{fuch96,uma97}.

Pions are produced mainly by resonance decays, and this
pion production process is implemented in the model in the
following way. Nucleons while propagating shall collide 
stochastically if the
distance between the centroids of the two Gaussian wavepackets 
is less than $d_{\rm min}=\sqrt {\sigma_{\rm tot}({\sqrt s})/\pi }$.
This collision process is implemented using Monte Carlo methods and
the effect of the Pauli exclusion principle is duly taken into account.
For the inelastic 
nucleon-nucleon channels we include the $\Delta(1232)$ as well as 
the $N^{*}(1440)$ resonance. In the intermediate energy range the 
resonance production is dominated by the $\Delta$, however, the 
$N^{*}$ yet gives non-negligible contributions to the high energetic 
pion yield \cite{metag93}. The 
resonances as well as the pions originating from their decay are 
explicitly treated, i.e. in a non-perturbative way and all relevant 
channels are taken into account. 
In particular we include the resonance production and rescattering 
by inelastic NN collisions, the one-pion decay of $\Delta$ and 
$N^{*}$ and the two-pion decay of the $N^{*}$ and one-pion 
reabsorption processes. ( For details see Ref. \cite{fuch96}.) 
The isospin dependence of the pion production cross sections is 
duly included in the model. Thus one is able to understand
the observed charged dependence given by 
$N(\pi^-) > N(\pi^0) > N(\pi^+)$. 
For the Au+Au system, we obtain the ratios $\pi^-:\pi^0:\pi^+ \ = \ 
1.37:1:0.71$ for the respective multiplicities, which then gives
$\pi^-/\pi^+=1.95$.
Finally, the pions thus produced are guided essentially 
by the Coulomb interaction between the pions and baryons.

It may be said here that the pion propagation can, in principle, 
be also affected by the contribution arising
from the effect of nuclear medium, i.e. from
the collective $\Delta N^{-1}$ and $NN^{-1}$
excitations,  on the pion dispersion relation.
However, it is shown in a recent study \cite{fuch96} that 
only weak corrections, i.e. a small in-medium pion-nucleus
potential, are required in order to obtain a realistic description
of pionic observables.
 In addition, also in other works \cite{eheh93,xion93} it was found
that the high energy part of the pion spectrum is not much 
affected by these medium corrections. 
A more sophisticated treatment of collective pionic 
excitations within the transport approach was proposed
in Ref. \cite{helg95}, which was subsequently applied to
heavy ion collisions in Ref. \cite{helg97}. However, as
discussed in Ref. \cite{fuchs97}, it is rather questionable
if the simple $\Delta N^{-1}$ ( and $NN^{-1}$)
model yields a reliable pion dispersion relation.
Further, the results of Ref. \cite{helg97} indicate
that the influence of collective excitations on
pionic observables is small and tends to enhance the
pion yields with respect to the standard approaches 
and the experimental findings.
Moreover, in this exploratory study we wish to 
concentrate on the role of the IDMF in pion production.
Hence, we omit the medium effects of 
the charge dependence of pion production in the present work.

Now, coming to the baryon propagation part, the transport models
like QMD normally propagate the nucleons and resonances in a 
density and momentum dependent nuclear mean field and as well
include the Coulomb interaction among the charged particles.
In the present study we consider in addition to the above
stated components a isospin dependent contribution to the
total mean field. Thus, one has,
\begin{equation}
V_i = V_{\rm symm}(\rho, |{\bf p}|) + V_{\rm Coul}^i + 
V_{\rm asym}^i(\rho, I)
\label{mf}
\end{equation}
where $i$ stands for any one of the nucleons or resonances.
The isospin dependent potential $V_{\rm asym}^i$ is determined 
following Ref. \cite{li97} as
\begin{equation}
V_{\rm asym }^q = {\partial \over \partial \rho_q}
\epsilon_{\rm asym} \quad ; \quad q=n,p
\label{vasym}
\end{equation}
where $\epsilon_{\rm asym} = \rho S_{\rm potn}(\rho_o) 
F(\rho / \rho_o ) $ and $S_{\rm potn} (\rho_o )$ is the
potential part of the nuclear symmetry energy at nuclear
matter saturation density ($\rho_{\rm sat} = 0.16$ fm$^{-3}$). 
The density dependence of the IDMF is taken to be
$F(\rho /\rho_o ) =  {(\rho / \rho_o )}^{\gamma}$.
The strength of the density dependence varies widely
among the theoretical studies 
\cite{siem70,chin77,horo87,prak88,muth87}. 
For our present qualitative
investigation, we choose two particular values, $\gamma=0.3$ and
$\gamma=2.0$.
Having obtained the IDMF for neutrons and protons, we now
express $V_{\rm asym}$ for the respective resonances as: 
$V_{\rm asym}(n^*) = V_{\rm asym}(\Delta^0 ) = V_{\rm asym}^n$, 
$V_{\rm asym}(p^*) = V_{\rm asym}(\Delta^+ ) = V_{\rm asym}^p$, 
$V_{\rm asym}(\Delta^{++}) = 2V_{\rm asym}^p-V_{\rm asym}^n$ 
and $V_{\rm asym}(\Delta^{-}) = 2V_{\rm asym}^n-V_{\rm asym}^p$. 
Thus, both, nucleons and resonances propagate in an isospin
dependent mean field.

Consequently, there shall appear -- with the other conditions 
being the same -- a smaller number of $n$--$n$ and $n$--$p$ collisions, 
and a slightly enhanced number of $p$--$p$ collisions. 
Since $\pi^-$ are predominantly created in
$n$--$n$ collisions via $\Delta^-$ and $\Delta^0$ decays, the 
number of $\pi^-$ ($N_{\pi^-}$) is expected to decrease. 
On the other hand, as $\pi^+$ are mainly created in
$p$--$p$ collisions via $\Delta^{++}$ and $\Delta^+$ decays, 
$N_{\pi^+}$ may slightly increase. Therefore, the charged
pion ratio $\pi^-/\pi^+$ may effectively decrease.
But, we need to take into account the effect of re-absorption
process as well, which sort of counterbalances the effect due to
the IDMF on pion production.
In addition, in the case of $\pi^+$, both $n$--$p$ and $p$--$p$ collisions
contribute to $N_{\pi^+}$ via $\Delta^+$ decay. As $N_{<np>} > N_{<pp>}$ 
there may as well be a slight reduction in $N_{\pi^+}$ at the end
of the reaction.
The final outcome of the interplay of these effects 
in regard to pionic observables is
discussed in the following section.

\section{Results and Discussions }

In this section we discuss the results as obtained from simulations
made including an isospin dependent mean field in the baryon
propagation part. All the calculations pertain to the $Au+Au$ 
reaction at $1$ A.GeV and use the standard momentum dependent 
Skyrme force which corresponds to a soft equation 
of state, hereafter referred to as SMD force.

As expected, we find that the average number of collisions are
modified due to the presence of a IDMF. It is found that 
average number of $n$--$n$ collisions, $N_{<nn>}$, decrease by
about $10\%$, while $N_{<pp>}$ increase by about $5\%$.
As a result, it can be seen from Fig. 1 that, the total pion
multiplicity $N_{\pi}$ obtained for zero impact parameter, 
 $b=0$, decreases by about $20\%$ at midrapidity.  
Moreover, though one sees that the reduction in 
$N_{\pi}$ slightly increases with increase in the value of parameter
 $\gamma$, which signifies the density dependence of the symmetry energy; 
the dependence of $N_{\pi}$ on $\gamma$ is not so prominent
at high incident energies considered here.
In addition, it may be said here that though the total pion number
at $b=0$ decreases from $\sim 58$ to $\sim 50$, the theoretical 
results obtained for charged pion multiplicity 
are still overestimated \cite{pelt97,uma97} 
as compared to the experimental data.

It is known \cite{bass94} that the high energy pions belong mostly to the
early, compressed phase of the nuclear reaction. Hence, these
pions undergo relatively less scattering as compared to the
low energy pions, and thereby, one expects high energy
pions to be less affected by the isospin dependence of the mean field. 
This aspect is illustrated 
in Fig. 2, where the pion transverse momentum $p_t$ distribution 
calculated at midrapidity with $b=0$ is shown. A substantial
reduction over the low$-p_t$ part due to the IDMF can be noted. 
In contrast, the high$-p_t$ part remains more or less the same. 
Further, a similar suppression of low$-p_t$ pions is also 
found in the case of non-central collisions. 
It maybe recalled here that transport models like QMD overestimate 
the low energy part of pion spectra. Now, with inclusion of the IDMF, 
we expect that this overestimation of model calculations as 
compared to the data to be reduced. 
Toward this purpose, we performed simulations pertaining to the
minimum bias condition for a particular value of $\gamma=2$. 
Results obtained for the charged pion spectra
are compared in Figs. 3 and 4 to the FOPI data \cite{pelt97}.
It is satisfying to note that results obtained with the IDMF are 
closer to the FOPI data than those obtained with $V_{\rm asym}=0$, 
particularly over the low$-p_t$ part. In addition, the effect of 
the IDMF is more significant on the $\pi^-$ spectrum than on $\pi^+$. 
Thus, for a quantitative description of pion yields and spectra, we
need to include the isospin dependence of the mean field.

Another quantity which is sensitive to the isospin dependence
is the charged pion ratio $\pi^-/\pi^+$.
The study of the charged pion ratio is of importance and current 
interest since one can, in principle, extract information 
regarding the size of the participant zone during 
 its compression and expansion stages.
In particular, the observed energy dependence of $\pi^-/\pi^+$
$-$ that is, the value of this ratio decreases from about 3.0 
to 0.9 as pion kinetic energy increases from $\sim 50$ MeV to
$\sim 350$ MeV, and remains almost constant at higher energies$-$ 
is shown to be due to the Coulomb potential \cite{uma97,bass95,teis97a}, 
which is opposite in nature for $\pi^+$ and $\pi^-$.
With the inclusion of a IDMF we expected the ratio to decrease, 
particularly over the low energy region. 
However, we found that in practice the decrease is not
substantial enough to illustrate the role of IDMF.
This may be due to, firstly, that there is a net reduction in 
low$-p_t$ $\pi^{+'}s$ as these positively charged pions are also
produced in $n-p$ collisions, and secondly, that the low$-p_t$
pions undergo more production-reabsorption cycles
before they freeze out which may wash out the isospin effects.

However, the effect of the IDMF on $\pi^-/\pi^+$ is more clearly 
seen when the charged pion ratio is analysed as a function of the
normalized center--of--mass rapidity $Y^0$.
Results obtained with $b=0$ are shown in Fig. 5. The error bars 
quote the statistical uncertainty of the calculations which 
is due to the event number. Quite interestingly, 
similar to the energy dependence, there appears a systematic 
$Y^0-$dependence of the $\pi^-/\pi^+$ ratio where we obtain 
in the case of no asymmetry dependence, i.e. $V_{\rm asym}=0$, 
$\pi^-/\pi^+ > 1.95$ over the midrapidity region and
the ratio decreases with increasing $\mid Y^0 \mid$. 
This $Y^0-$dependence is essentially due to the Coulomb interaction. 
This is also illustrated in the same figure, where one finds that,  
in the absence of the Coulomb interaction, the ratio remains
more or less constant, i.e. $\pi^-/\pi^+ \sim 1.8$ which is 
indicated by the horizontal line in Fig. 5. 
Furthermore, once the IDMF is included, the charged pion ratio 
gets significantly suppressed over the midrapidity region while the 
distribution remains more or less the same beyond the midrapidity
region. With an increase in the value of $\gamma$, Eq. (\ref{vasym}), 
the amount of suppression slightly increases. 
In addition, there appears a double peaked structure, even at $b=0$, 
which is due to the isospin dependence of the nuclear mean field. 
Furthermore, this structure appears to be sensitive on the density 
dependence of the isospin dependent interaction. 
Thus, this particular observable clearly illustrates the 
effect due to Coulomb interaction and the IDMF.
A recent preliminary analysis \cite{pelte} of the 
FOPI data for 1 A.GeV $Au$ on $Au$ collisions 
indicates such a double peaked structure for semi-central 
as well as for most central collisions. In semi-central collisions 
this structure is probably due to an incomplete stopping and 
thermalization of the pions and can be qualitatively 
explained by the standard approach. To illustrate this effect we 
show in Fig. 6 the  $\pi^-/\pi^+$ ratio for a central (b=1) and 
a semi-central (b=6) collision with both calculations performed 
without an isospin dependence of the nuclear mean field. As already 
seen in Fig. 5 there exists only one single peak for b=1. Thus the 
reduction of the $\pi^-/\pi^+$ ratio at midrapidity in central 
collisions can probably be regarded as 
a signature for the importance of the IDMF in the pion 
production mechanism at SIS energies.

\section{Summary }
To summarize, we have made an exploratory study on the importance
of a isospin dependent nuclear mean field (IDMF) concerning the pion 
production mechanism in intermediate energy heavy ion reactions. 
It is found that due to the different isospin charges of nucleons
and resonances, the average number of collisions gets modified 
and hence, the pion yield.
In particular, the low energy pions get significantly suppressed
rendering a better agreement of calculated pion spectra with the
FOPI data. We have also analysed the rapidity distribution of 
the charged pion ratio where due to the effect of the IDMF, one finds
a substantial reduction of the $\pi^-/\pi^+$ ratio over the midrapidity
region in central collisions. This particular observable seems to
illustrate clearly the effects due to Coulomb force and the IDMF.
Therefore, our present study strongly suggests that at the
quantitative level, it is quite important to take into
account the isospin dependence of the nuclear mean field.
\begin{acknowledgments}
We thank D. Pelte for fruitful discussion. 
\end{acknowledgments}

\newpage
\newpage
\begin{figure}
\begin{center}
\leavevmode
\epsfxsize = 15cm
\epsffile[100 130 440 470]{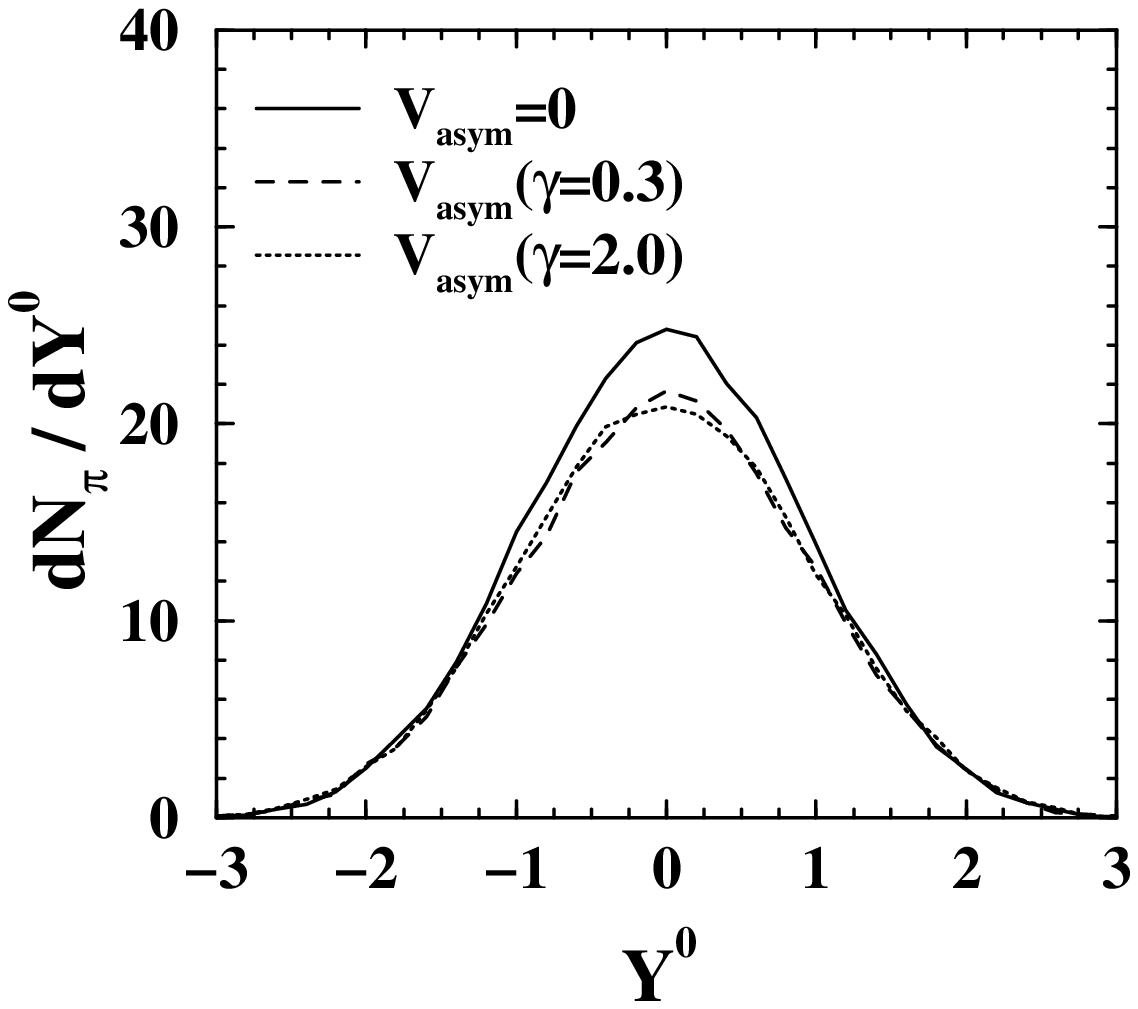}
\end{center}
\caption{
The total pion multiplicity $N_{\pi}$ in a central ($b=0$)
$Au + Au$ reaction at 1 A.GeV is shown as a function 
of the normalised rapidity $Y^0=
{(Y_{\pi}/Y_{\rm proj})}_{\rm cm}$.    
The calculations are performed with (dashed, dotted) 
and without (solid) an isospin dependent nuclear mean field.
}
\label{fig1}
\end{figure}
\begin{figure}
\begin{center}
\leavevmode
\epsfxsize = 15cm
\epsffile[90 130 445 510]{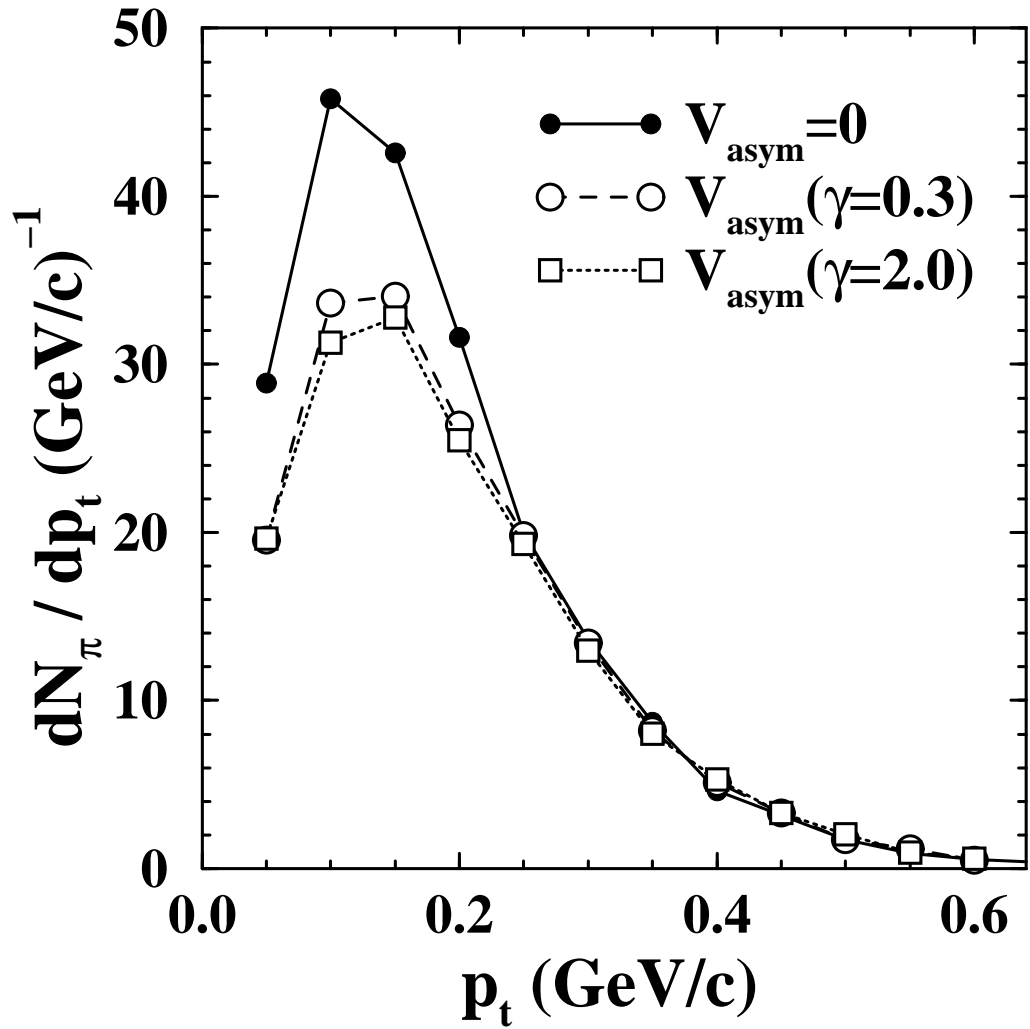}
\end{center}
\caption{
Pion transverse momentum $p_t$ distribution calculated 
at midrapidity $-0.2 \le Y^0 \le 0.2$ for the same reaction 
as in Fig. 1.
}
\label{fig2}
\end{figure}
\begin{figure}
\begin{center}
\leavevmode
\epsfxsize = 15cm
\epsffile[0 50 480 470]{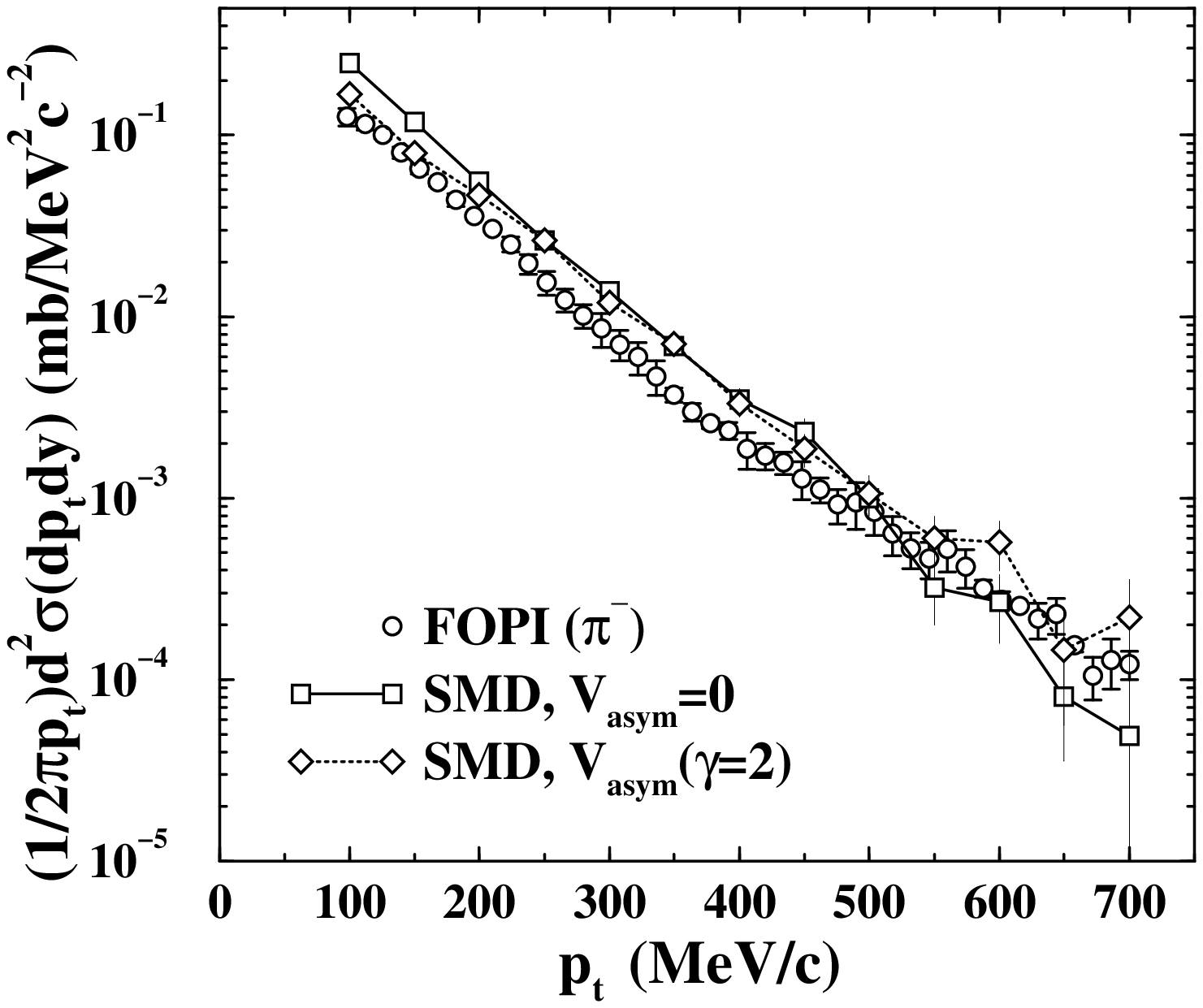}
\end{center}
\caption{
The $\pi^-$ $p_t$ spectrum in a 1 A.Gev $Au+Au$ reaction 
obtained at midrapidity 
under {\it minimum bias} condition is compared to the 
FOPI data of Ref. \protect\cite{pelt97}.
}
\label{fig3}
\end{figure}
\begin{figure}
\begin{center}
\leavevmode
\epsfxsize = 15cm
\epsffile[0 50 480 470]{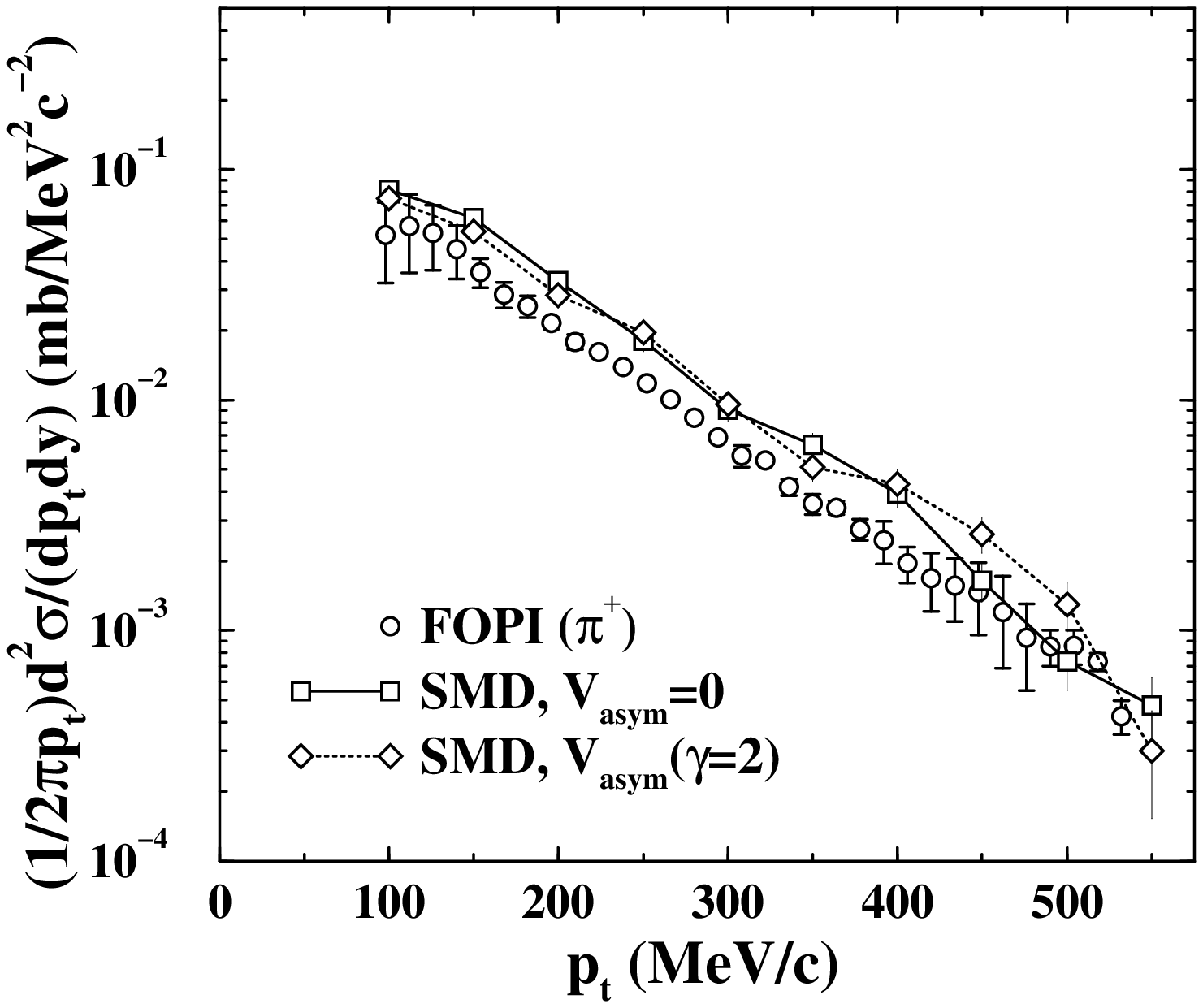}
\end{center}
\caption{
The $\pi^+$ $p_t$ spectrum in a 1 A.GeV $Au+Au$ reaction 
obtained at midrapidity 
under {\it minimum bias} condition is compared to the 
FOPI data of Ref. \protect\cite{pelt97}.
}
\label{fig4}
\end{figure}
\begin{figure}
\begin{center}
\leavevmode
\epsfxsize = 15cm
\epsffile[60 130 450 510]{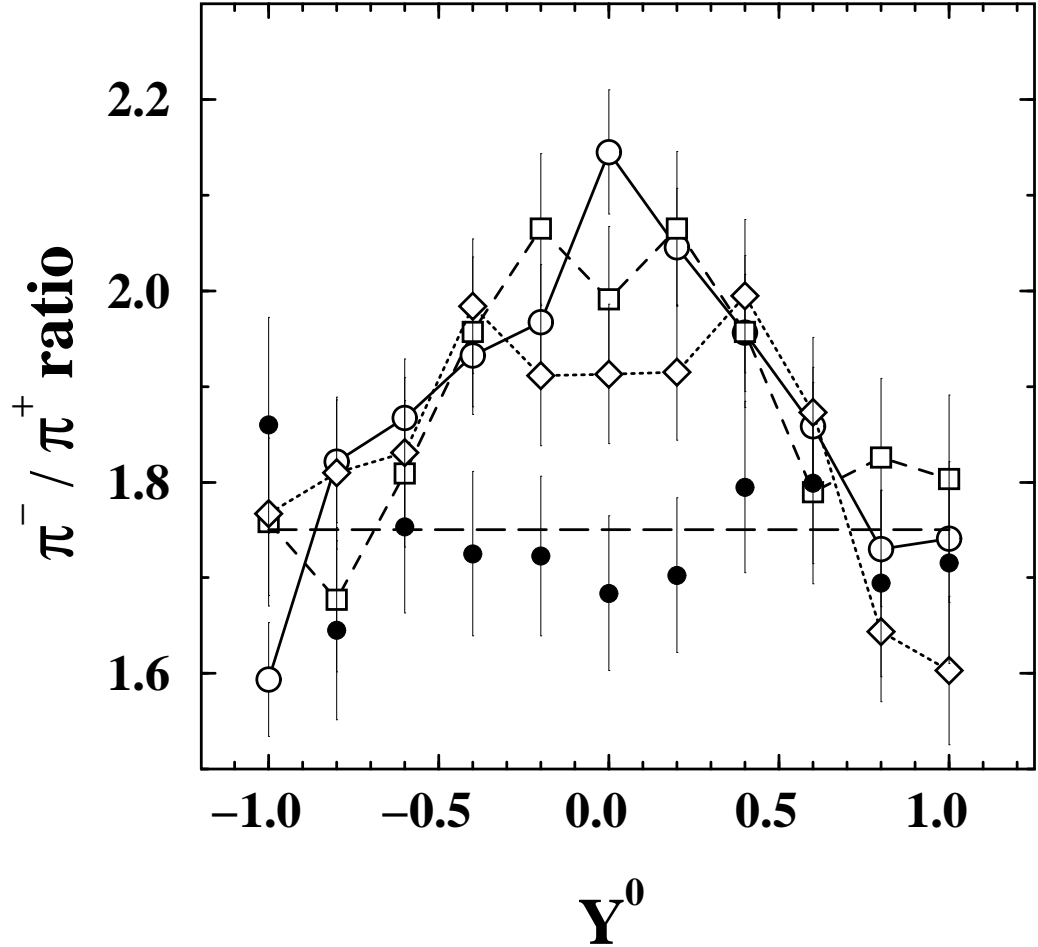}
\end{center}
\caption{
Dependence of the charged pion ratio on the normalized rapidity
calculated in a central ($b=0$) 1 A.GeV $Au + Au$ reaction.
The calculations are performed without (open circles) and 
including the isospin dependence of nuclear 
mean field. Here squares refer to the IDMF given by the parameter 
$\gamma = 0.3$ and diamonds to $\gamma = 2.0$. The filled circles 
correspond to a calculation without
both, the isospin dependence and the pion-baryon 
Coulomb interaction. 
The horizontal line refers to the average value of the charged
pion ratio obtained in this calculation.
}
\label{fig5}
\end{figure}
\begin{figure}
\begin{center}
\leavevmode
\epsfxsize = 15cm
\epsffile[60 130 450 510]{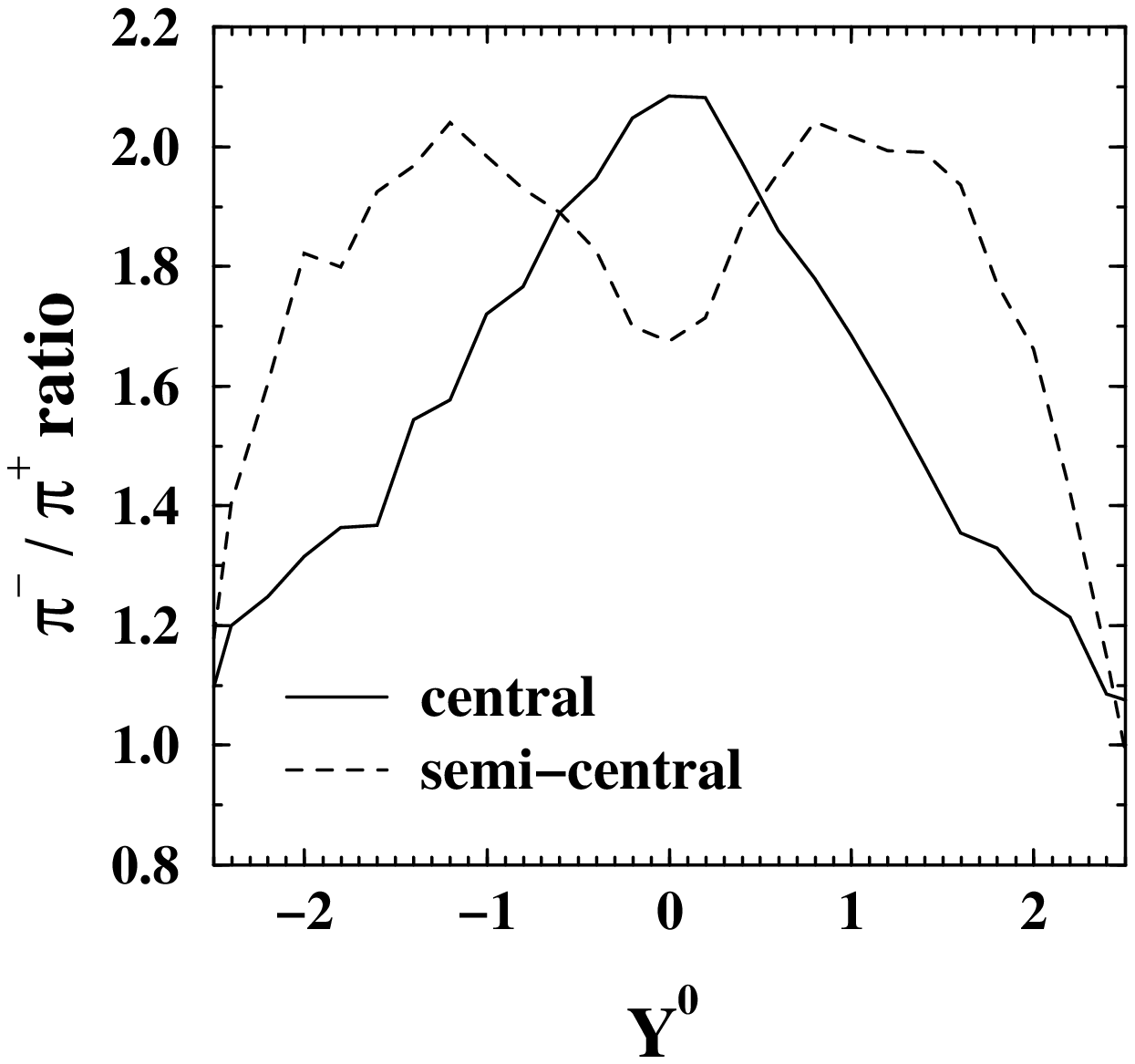}
\end{center}
\caption{
Centrality dependence of the charged pion ratio on 
the normalized rapidity in a 1 A.GeV $Au + Au$ reaction.
The calculations for the central (b=1 fm) and the semi-central 
(b=6 fm) reaction are performed with the standard SMD force, i.e. 
without an isospin dependence of nuclear 
mean field. 
}
\label{fig6}
\end{figure}

\begin{thebibliography}{99} 

\bibitem{pelt97}
D. Pelte and the FOPI Colloboration, Z. Phys. A 357 (1997) 215.

\bibitem{bril93}
D. Brill and the KaoS Colloboration, Phys. Rev. Lett. 71 (1993) 336. 

\bibitem{munt95}
C. M\"untz and the KaoS Collaboration, 
Z. Phys. A 352 (1995) 175.

\bibitem{naga81}
S. Nagamiya, M.C. Lemarie, E. Moeller, S. Schnetzer, G. Shapiro, 
H. Steiner, and I. Tanihata, Phys. Rev. C 24 (1981) 971.

\bibitem{broc84}
R. Brockmann {\it et al.}, Phys. Rev. Lett. 53 (1984) 2012.

\bibitem{kint97}
J. C. Kintner {\it et al. }, Phys. Rev. Lett. 78 (1997) 4165

\bibitem{teis97}
S. Teis, W. Cassing, M. Effenberger, A. Hombach, U. Mosel, and
Gy. Wolf, Z. Phys. A 356 (1997) 421.

\bibitem{uma97}
V. S. Uma Maheswari, C. Fuchs, A. Faessler, L. Sehn, D. Kosov, Z.S. Wang,
NUCL-TH/9706004.

\bibitem{pelt97a}
D. Pelte and the FOPI colloboration, NUCL-EX/9704009, Z. Phys. A in press.

\bibitem{stoc86}
R. Stock, Phys. Rep. 135 (1986) 259.

\bibitem{verw82}
B.J. VerWest and R.A. Arndt, Phys. Rev. C 25 (1982) 1979.

\bibitem{eric88}
T. Ericson and W. Weise, Pions and Nuclei, 
Carendon Press Oxford 1988.

\bibitem{li96}
B.A. Li, Z.Z. Ren, C.M. Ko and S.J. Yennello, Phys. Rev. Lett. 76 (1996) 4492.

\bibitem{li97}
B.A. Li, C.M. Ko and Z.Z. Ren, NUCL-TH/9701048, 
Phys. Rev. Lett. in press.

\bibitem{aich91}
J. Aichelin, Phys. Rep. 202 (1991) 233.

\bibitem{metag93}
V. Metag, Prog. Part. Nucl. Phys. 30 (1993) 75.

\bibitem{fuch96}
C. Fuchs, L. Sehn, E. Lehmann, J. Zipprich and A. Faessler, 
Phys. Rev. C 55 (1997) 411.

\bibitem{eheh93}
W. Ehehalt, W. Cassing, A. Engel, 
U. Mosel and Gy. Wolf, Phys. Lett. B 298 (1993) 31. 

\bibitem{xion93}
L. Xiong, C.M. Ko and V. Koch, Phys. Rev. C 47 (1993) 788. 

\bibitem{helg95}
J. Helgesson and J. Randrup, Ann. Phys. (N.Y.) 244 (1995) 12; \\
Nucl. Phys. A 597 (1996) 672.

\bibitem{helg97}
J. Helgesson and J. Randrup, NUCL-TH/9705022, Phys. Lett. B in press.

\bibitem{fuchs97}
C. Fuchs, L. Sehn, E. Lehmann, J. Zipprich and A. Faessler, Phys. Rev. C 
56 (1997) 1189.

\bibitem{siem70}
P.J. Siemens, Nucl. Phys. A 141 (1970) 225

\bibitem{chin77}
S.A. Chin, Ann. Phys. (N.Y.) 108 (1977) 301.

\bibitem{horo87}
C.J. Horowitz and B.D. Serot, Nucl. Phys. A 464 (1987) 613

\bibitem{prak88}
M. Prakash, T.L. Ainsworth and J.M. Lattimer, Phys. Rev. Lett. 61 (1988) 2518.

\bibitem{muth87}
H. M\"uther, M. Prakash and T.L. Ainsworth, Phys. Lett. B199 (1987) 469.

\bibitem{bass94}
S.A. Bass, C. Hartnack, 
H. St\"ocker and W. Greiner, Phys. Rev. C 50 (1994) 2167.

\bibitem{bass95}
S.A. Bass, C. Hartnack, 
H. St\"ocker and W. Greiner, Phys. Rev. C 51 (1995) 3343.

\bibitem{teis97a}
S. Teis, W. Cassing, M. Effenberger, A. Hombach, U. Mosel, and
Gy. Wolf, NUCL-TH/9701057.

\bibitem{pelte}
D. Pelte, private communication.

\end{thebibliography}
\end{document}